\documentclass[printer]{aa} 
\usepackage{amsmath}

\usepackage{graphicx}
\usepackage{txfonts}
\usepackage{natbib}

\usepackage{color}
\newcommand{\be}{\begin{eqnarray}}
\newcommand{\ee}{\end{eqnarray}}

\newcommand {\nbodypp}{\textsc{\mbox{nbody6\raise.4ex\hbox{\tiny++}}}}

\newcommand {\Msun} {\mbox{M$_{\odot}$}}

\begin{document}

\title{Formation and dissolution of leaky clusters}
\author{S. Pfalzner}
\institute{Max-Planck-Institut f\"ur Radioastronomie, Auf dem H\"ugel 69, 53121 Bonn, Germany\\ \email{spfalzner@mpifr-bonn.mpg.de}}
\date{ }

  \abstract
   {Massive Galactic clusters ($> 1000$ \Msun) exhibit a clear correlation between cluster density, size and age and can be sorted in two categories, i.e. starburst and leaky clusters. The reason for the existance of two types of massive clusters is an open question. However, the answer is probably connected to a different formation histories of the two types.}
   {In this study we concentrate onleaky clusters only and investigate possible formation scenarios and gas expulsion phase.  }
   {This is done by using existing observational data and numerical  results of embedded cluster properties. }
   {Assuming that a clear correlation between cluster density, size and age exists, it is shown that the density-radius development over time forembedded clusters  can be approximated by $\rho \approx 100*r ^{-1.3}$\Msun  pc$^{-3}$. The consequences for the star formation process in leaky clusters are discussed and found to favour an inside-out star formation scenario with an initially low but later accelerated star formation rate. It is shown how the leaky clusters form in a unique sequential manner and that rapid gas expulsion is responsible for the  80-90\% mass loss over the next 20 Myr. }

   \keywords{star cluster dynamics -- star formation --gas expulsion
               }

\maketitle

\section{Introduction}

\bigskip
\bigskip
\bigskip
\noindent
Most stars form in dense clusters embedded within giant molecular cloud clumps\citep{lada:03,pudritz:02}. Given the relatively high number of clusters younger than 10 Myr compared to the much lower number of older clusters in equivalent age spans,  the majority of clusters must rapidly disolve soon after their formation. As no clear correlation between
cluster mass and age were found, until recently the general view has been that in the Milky Way, star formation results in  a multitude of cluster with hardly any constraints on stellar content, size or density. However, at least for clusters more massive than 1000 \Msun\  this view had to be revised. \cite{pfalzner:09} showed that these massive clusters follow two separate sequences. Although these are difficult to distinguish in the mass-age plane they become immediately apparent in density-radius space. In both cases the expulsion of gas from the cluster via stellar winds, ionisation and supernovae explosions\citep{hills:80,goodwin:06,bastian:08,baumgardt:07} leads to an expansion
 (see Fig. 1 here and more details in \cite{pfalzner:09} Fig.~2).

The reason of this bi-modal development is currently not understood. One might think that a difference in cluster mass might cause it. However, Fig. 1 shows that at an early age ($<$4Myr) both types of clusters have nearly the same mass but the cluster radii in the two groups (marked by red diamonds) differ by more than a factor 10. An additional difference is that the mass of starburst clusters remains more or less constant whereas that of leaky clusters reduces to about 10\% - 20\% of its initial value. A hint to the origin of the two groups of massive clusters is the obvious difference of the {\em location} of the clusters - all starburst clusters are located in regions of high stellar density - near the galactic center or in the spiral arms - whereas the leaky clusters are situated in lower density areas. Clearly the answer to the question of why massive clusters exist in two categories must lie in their formation process. In this paper we concentrate on the formation and development of leaky clusters  which start out with masses of the order of 10$^4$ \Msun\ and loose 80\%-90\% of this mass over the next 20 Myr after gas expulsion.

\begin{figure}
\includegraphics[width=84mm]{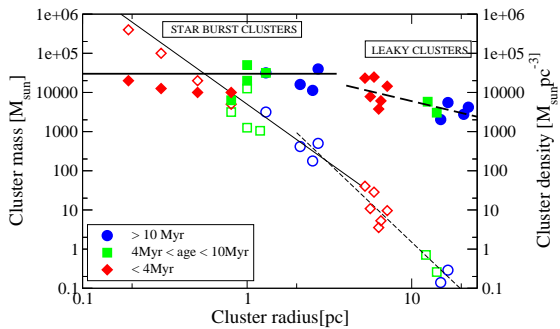}
\caption{Cluster mass as a function of cluster size (full symbols) for exposed clusters more massive than 10$^3$\Msun. The diamond symbols represent clusters younger than 4 Myr, the   squares clusters in the age range 4Myr $<t_c<$ 10 Myr and the circles clusters with ages 10Myr $<t_c<$ 20 Myr. The thick drawn line shows that the cluster mass remains more or less constant for the starburst cluster sequence, whereas the leaky cluster mass rapidly declines with cluster age (thick dashed line).The values were taken from \cite{figer:08},  \cite{wolff:07} and \cite{borissova}. In addition the relation between the cluster density and cluster radius are shown as in \cite{pfalzner:09} by empty symbols.}
\end{figure}

Leaky clusters form as a  bound entity which dissolves to a large degree after gas expulsion.  The exphasis of this study lies on the formation process of leaky clusters. The precursors to leaky clusters must be found among the still embedded clusters. But the question is which embedded clusters develop into leaky clusters and how does this development proceed. 

One possibilty would be that leaky clusters form by merging of sub-clusters like recently suggested by several authors for less massive cluster (see, for example, K\"upper et al. 2011 and references therein). However, to form a leaky cluster this would require that several clusters like the Orion Nebula cluster (ONC) with $>$ 1000 stars or hundreds of small-N clusters would have to form in close vicinity and merge within $<$ 1Myr. So far there exist no observational indications for such a formation process for leaky clusters. Here we stick to the more conservative picture of a leaky cluster forming as a single entity. 

In the following existing data of embedded cluster properties are used to constrain the conditions of leaky cluster formation and the gas expulsion phase. The idea is that a better understanding of the formation process and later development of leaky clusters will eventually answer the open question why leaky and starburst clusters form and develop differently.



\section{Method}

This work uses available observational data of the properties of embedded cluster by \cite{lada:03} to propose a possible formation history for leaky clusters. In principle one could deduce the star formation history of massive clusters by looking at the age distribution within single such clusters. However, there are two problems with this approach: i) leaky clusters show
considerable mass loss within the first 10 Myr of their developement and ii) it is not clear how long the process of star formation actually takes in such clusters - it could be coeval or take up to $\sim$10 Myr.  

In young embedded clusters like the ONC, wide spreads in colour-magnitude diagrams (CMD) have been observed, which have previously been interpreted as age spreads of up to 10 Myr in association members\citep{palla:00}. However, other authors have
 suggested that the observed CMD spreads are due to a mixture of photometric errors, variability in pre-MS stars \citep{herbst:94}, binarity \citep{preibisch:99}, episodic accretion phases\citep{baraffe:09} or a spread in line-of-sight distances or extinctions. Currently age spreads of 1-3 Myrs are regarded as realistic for the ONC. 
However, the age of Cyg OB2 was originally estimated as 1-3 Myr considering only OB stars. A concentration of 5-7 Myr old A stars was found just to the south of Cyg OB2 (Drew et al. 2008) while 
an unclustered spread of evolved stars up to 10 Myrs old was identified over a wider area\citep{comeron:08}. \cite{wright:10} favour the view that the present-day OB stars are products of the latest phase of star formation, but reason that at least one more epoch of star formation took place beforehand. In summary, the observations of the stellar ages in single star clusters give a rather unclear picture of the duration and the dynamics of the star formation process.

%

%
%
%

So rather than looking at the star formation history in one single cluster one could look at  clusters of different age and construct a star formation history from this sequence.
Mature leaky clusters consist of thousands of stars, but nevertheless they must have started out in the embedded phase containing just a few dozen stars, aquiring additional members via star formation.  The straightforward approach would be to determine the ages of the embedded clusters and sort them temporally to obtain the formation history of leaky clusters. Unfortunately this is hindered  by the fact that age determination in the embedded phase is even more complicated than in the early exposed phases due to the high uncertainty of the pre-main sequence (PMS) stellar evolution models. So although embedded clusters are usually attributed ages in the 0-2 Myr range, the error bar is generally of the same order. As a consequence ordering embedded clusters according to their age would be extremely error prone. 

 For the exposed phase an important consequence of the sequential nature of the leaky cluster development in the cluster radius-age space is that in this case the radius of a cluster can  be used to determine its age (see Fig.3 in \citet{pfalzner:09}).  For leaky clusters  a fit to the observational data gives a correlation between the average cluster radius $r_{cl}$ and its age $t_{cl}$ of the form:  
\be 
      t_{cl}/\mbox{Myr} &\sim& 0.15 (r_{cl}/\mbox{pc})^{3/2}
. \ee 
%

One way out of the dilemma of determining a formation history for leaky clusters is the assumption that, similar to the exposed phase of the leaky cluster development, there exists a similar correlation between the cluster radius and its age in the embedded phase.
In other words, we postulate a relation between cluster membership i.e. of the number of stars already formed, and age  in the  embedded phase for star clusters that eventually develop into leaky clusters. This is the simplest assumption we can make: it implies that star formation takes place in these clusters. 

Now implicitly the same underlying reasoning as for exposed clusters is used: Since each of the exposed leaky clusters has the same developmental history, it seems reasonable to assume that their precursors in the embedded phase have as well a common history.  However, not every embedded cluster will develop into a massive leaky clusters. Only if there is sufficient matter in the surrounding cloud this developmental path is open.  Clusters with too low mass content in their surrounding stop
forming stars as soon as they run out of material - well before they have reached masses in excess of 10$^3$ \Msun.  Most likely these starved clusters quickly disperse so that their surface density more or less immediately drops below the detection line.

%




Back to the formation of leaky clusters - 
in order to investigate the consequences of the assumption of a common history of the precursors of the leaky clusters, we use the data of embedded cluster properties by Lada \& Lada (2003). They selected these clusters within 2kpc from the sun
in such a way that the group must have a stellar-mass volume density sufficiently
large to render it stable against tidal disruption by the Galaxy (i.e., $\rho_∗>$0.1 \Msun
pc$^{-3}$) and by passing interstellar clouds (i.e., $\rho_∗>$1 \Msun pc$^{-3}$). In addition, the group should contain a sufficient number of members to ensure that its evaporation time is greater than the typical lifetime of Galactic open clusters.  It means that possibly existing clusters with densities $< 5 \Msun (r_{cl}/pc)^{-3}$ are not included in this sample given in
 \cite{lada:03}(see dashed line in Fig.2).

\begin{figure}
\includegraphics[width=84mm]{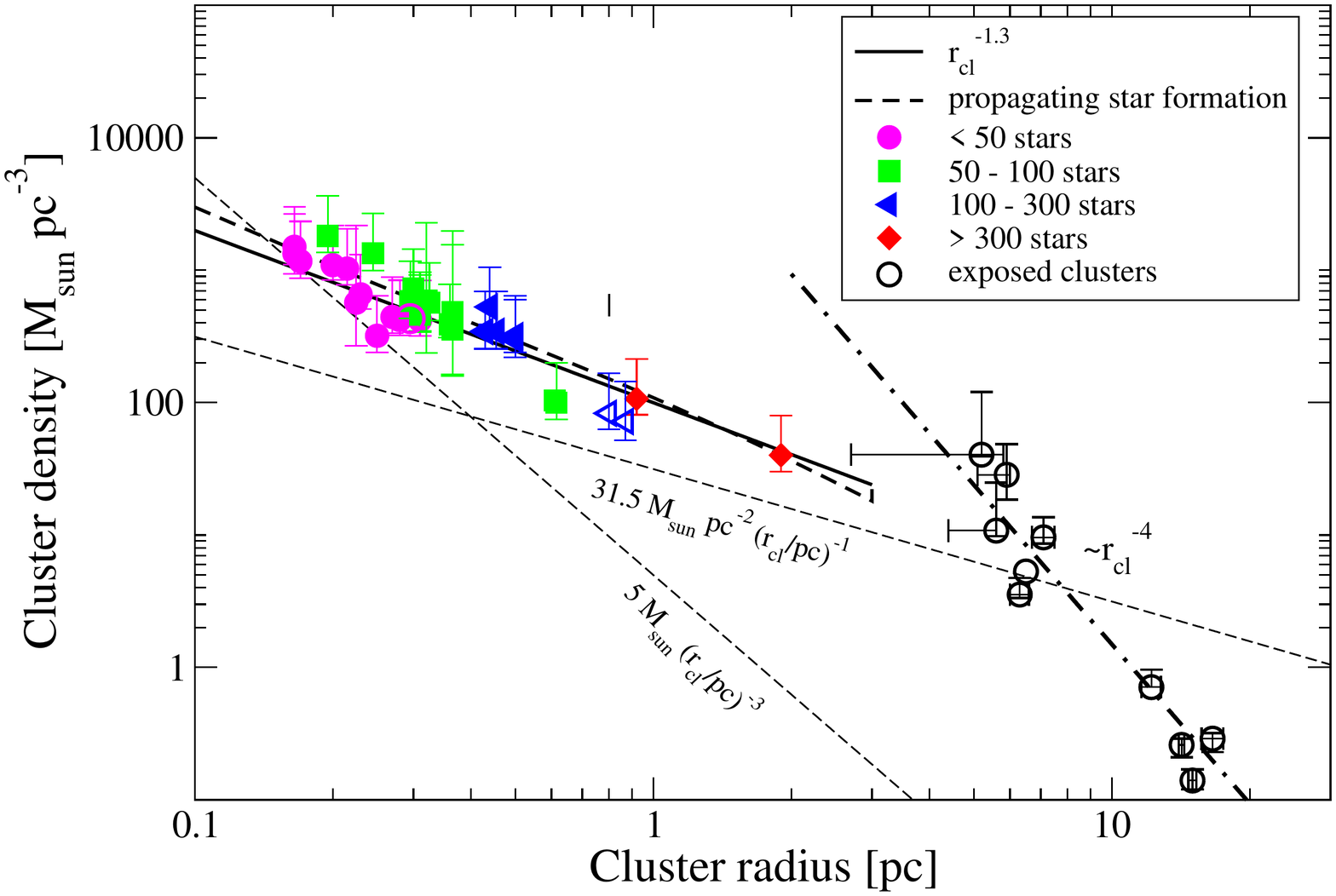}
\caption{Cluster density as a function of cluster size for the embedded clusters listed in
\cite{lada:03} (coloured symbols) and the leaky clusters from Pfalzner 2009 (black symbols).
The open coloured symbols for the embedded clusters indicate the values for  IRAS 06155+2319, GEM1, NGC 2281 and GEM4. The selection criteria by \cite{lada:03} imply that clusters with densities $< 5 \Msun (r_{cl}/pc)^{-3}$ are not included. In addition clusters with densities $< 31.5 \Msun pc ^ {-2} (r_{cl}/pc)^{-1}$ can not be detected against the field of background stars.
}
\label{fig:embedded_all}
\end{figure}

\cite{carpenter:00} state that for their sample  the observational density limits imply that clusters of density $<$ 31.5 \Msun pc$^{-2} (r_{cl}/$pc)$^{-1}$ are too tenuous to be detected against the background of the field. If this limit also applies to the \cite{lada:03} data
sample (see dashed line in Fig. 3.), the absence of lower density embedded clusters  might be an artifact of observational limitations.  On the other hand, this limit might also be too stringent - the exposed leaky clusters with sizes $>$10pc can be detected despite their density being relatively low compared to the background density.

\section{Results}

\subsection{Density development in the embedded phase}

Even the most massive  leaky cluster must have started out at some point with a small number of stars. Fig.~2 shows the lower part of Fig.~1 - the leaky cluster sequence - but
augmented by the data sample of embedded clusters from \cite{lada:03} with known sizes and masses. 
Because this data set is limited to embedded clusters within 2 kpc from the sun, its members are sufficiently far away from the Galactic center and the spiral arms, so it is most likely free from precursors to starburst clusters.
Fig. 2 implies that the precursors of leaky clusters form an extension of the leaky cluster sequence towards smaller radii ($<$ 3pc) at density of $\geq$ 100 \Msun pc $^{-3}$. However, the embedded clusters do not form a direct continuation of the slope of the main evolutionary sequence to lower radii, but the density development is flatter. This is because expansion due to gas expulsion has not started yet and star formation is still ongoing. 

In Fig. 2 and Fig.4 clusters with $<$50, 50-100, 100-300 and $>$300 members have been assigned to four different groups via distinct colours. The average radius of these cluster groups increases with the number of members. Following the time-sequence hypothesis - that age and radius are also correlated in the embedded phase - we may interpret Fig.~2 as follows : A cluster starts with an initial phase where 
a limited number of stars form ($<$100), the cluster radius increases by a factor 2-3 and the cluster becomes more diffuse. The relatively high number of embedded clusters is indicative of the slow star formation rate at this stage. 
Afterwards the star formation rate increases considerably and the cluster radius nearly grows by a factor of 10. 
During this phase the cluster density stays nearly constant or decreases only slightly. Whether these are actually 
two distinct phases or a continuous transition cannot be decided based on current data. However, considerable acceleration 
of star formation occurs at a cluster size of about 0.7-1pc when more than 100 members are formed.
In Fig. 2 the dependency of the cluster density $\rho$ on the cluster radius $r_{cl}$ 

\be \rho \approx 100 r_{cl}^{-1.3}\ee

provides only a rough clue to the development because the in the original data set not specified but presumably large errors in cluster radius do not justify a proper fit at this point.
The error in mass has been estimated as approximately a factor 2 by Lada \& Lada (2003), this error was taken into account in Fig.2 for determining the stellar densities.

Fig.2 shows that the average density in low N-clusters - which in the here presented picture corresponds to the youngest clusters - is often more than 10 times higher than that of the ONC. This means that in these early phases nearly all stars are exposed to interactions with neighbouring young forming stars which most likely influences star and planet formation processes.
Properties that might be altered by this high stellar density are, for example, the binary frequency, the density profile in the circumstellar discs, etc.
Although clusters like the ONC show a similar density in the central cluster regions to that of 
low-$N$ clusters, most stars are located in areas where the environment is less important.

The relative importance of these environmental effects on star and planet formation is still an
open question, however, the here provided knowledge of the density development during the star formation process allows to determine the encounter probability in such clusters better in future investigations. In the future this can show in how far the high stellar density influences star and planet formation processes in leaky cluster environments.

\begin{figure}
\includegraphics[width=84mm]{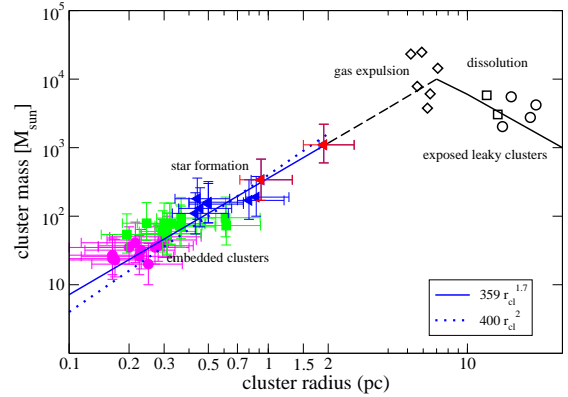}
\caption{ Mass as a function of radius for embedded clusters. The data points are taken from \cite{lada:03}
and the best fit for  $aR^b$ gives $a$=359$\pm$15 and $b$=1.71$\pm$0.07.
If the cluster mass is corrected to a lower mean stellar mass, this would change to
  $aR^b$ gives $a$=451$\pm$15 and $b$=1.71$\pm$0.07. 
However, given the large errors a $R^2$-relation could as well be considered. 
 The symbols are the same as depicted in Fig. 2.
\label{fig:mass_radius}}
\end{figure}

As mentioned before, not all embedded clusters included in Fig. 2 will necessarily develop into leaky clusters: Although the \cite{lada:03} sample probably excludes precursors to starburst clusters, it does not exclude precursors of lower mass clusters. Only if there is sufficient material in the surrounding cloud will the cluster continue to evolve towards the leaky cluster sequence. It is, for example, unlikely that IC348 at the leftward end of the embedded track in Fig. 2 will develop into a leaky cluster because its gas content is too low. For embedded clusters that develop into lower-mass exposed clusters the relation of Eq. 2 probably serves as an upper density limit. 

However, clusters that contain sufficient gas will eventually join up with the leaky cluster sequence when the gas expulsion has caused an expansion to $\sim$ 2-4pc. It is most likely that only embedded clusters as dense as those shown in Fig.2 can be the  percursors of the $<$ 4 Myr old leaky clusters. Lower density clusters might exist,  but those that are precursors to {\em leaky}  clusters will be at the high density limit (for a given radius) of this data set.


In Fig.2 the dependence of the cluster density on the cluster radius could as well be interpreted as an observational selection artifact, in the sense that many clusters show multiple peaks in their density. So if the low-mass clusters would all represent one such core and the higher mass clusters several cores with low density regions in between, one could obtain a similar diagram.  However, Fig.3 shows clearly that it is really a mass-radius relation. 

Plotting  the cluster mass of the \cite{lada:03} data as a function of the cluster radius (see Fig.~3) a relation of the form 
\be M_c =a r_{cl}^b \ee
with $a$=359$\pm$15 and $b$=1.71$\pm$0.07 is obtained.
It can be seen that the extension of Eq. 3 to larger cluster radii or equivalent in later times in the cluster development, meets with the leaky cluster track at the
point where these clusters become exposed. So Fig. 3 represents the formation, gas expulsion and dissolution phase of the leaky cluster sequence.  

In the following a rough estimate of the errors in cluster radii is attempted.
There are several sources that could lead to such an error. Due to the embedded nature not all stars are detected. If mass segregation is primordial, the higher likelihood of low-mass stars to be missed could lead to a too small detected radius. In addition, many of the embedded cluster are not spherically symmetric but are more extended in one direction. In Fig. 3 it was assumed that the error in cluster radius is approximately half its actual value.

\cite{adams:06} noted a related correlation between cluster radius and number of stars $N$, namely, $ r_{cl} \propto= N^{1/2}$. However, they interpret this correlation as a static property of embedded clusters and not as an outcome of temporal development of the cluster formation process. Nevertheless, with $M \propto N$ it would follow that $M_c \propto r_{cl}^2$ which considering the errors could as well fit the data (see Fig.3).

If the trend expressed by Eq. 3 is not an evolutionary sequence at all, then one has to explain why only certain mass-radius combinations seem to exist for young embedded clusters.
Selection effects could play a role, so it would be hard to detect clusters with  $M_c < $ 100 \Msun\ and $r_ {cl}< $ 2pc. However, it would be unlikely to miss clusters with, for example, $M_c >$ 500 \Msun\ and $r_ {cl}\sim$ 0.5pc, which are clearly not present in the sample.  One way out, would be a mass-size relation for cores (see, for example, Swift \& Williams 2008).  However, then the percursors of the
all clusters would have to form from thinly spread clusters with very low 
initial surface densities and be all below the detection limit. These clusters would then form stars
within a more or less fixed radius finally reaching such a density that they can be detected.
This explanation works very well for low-mass clusters. However, in such a picture the precursors of the leaky clusters should be embedded clusters with cloud masses $M_ {cloud} >$ 1000 \Msun\ containg thousands of stars within $r_ {cl}\sim$ 3-6pc. Such clusters should be detectable in the solar neighborhood. As this is not the case, the only explanation would be that they are either rare or very short-lived.
 
Therefore we prefer here the model of interpreting the mass-radius relation as a time-sequence. In this model the existence of a mass-radius relation naturally leads to such a form  as described by Eqs. 2 and 3 because these clusters still generate stars, so logically those with smaller populations must simply be younger. 

Eq. 3 probably directly describes the star formation process in leaky clusters since loss processes such as escapers play only a minor role. \cite{weidner:11} showed that in the embedded phase only about 15-20\% of the cluster mass is lost through escapers with little consequence to the cluster radius. If a direct relation between cluster radius and cluster age exists, Eq. 3 implies strongly {\em accelerated}  star formation.

\subsection{Star formation history}

What does the embedded cluster sequence  tell us about the star formation in a cluster environment? 
Two star formation scenarios would be possible:
i) either stars form only in the central high-density area, or ii) the star formation process itself progresses from the inside outwards.  We assume that the cluster size is a function of time of the form $r_{cl} \propto t^c$, where $c$ is a constant with  0$ \leq c \leq$ 1. Limiting our investigation to $c<1$ means an exclusion of acceleration cluster expansion.  Eq.~3 translates to $ M_c \propto t^{1.7 c}$.  
In the  case i)  the velocities would have to be supervirial so that as time progresses the cluster size expands. This could possibly be caused by an early gas expansion. However, this would require all clusters to have the same velocity dispersion. 

Case ii) is easier to reconcile with above findings. In a homogenoeus medium, one would expect  $ M_c \propto t^3$ if star formation just proceeded linearly from the centre outward. Since the observations indicate a $ M_c \propto t^{1.7 c}$-dependence,  this would either require  the star formation front to decelerate with time $r_{cl} \propto t^{0.6 c}$ or, the medium to become less dense. The latter fits what we expect from theory as well as observations - the gas density in the cluster forming clumps decreases from the centre outward.  

The gas density $\rho_g = M(r)/V$ in young forming clusters\citep{larsen:81,chandrasekhar:39} is either proportional to  $r^{-1}$  or $r^{-2}$.   Besides, power law density profiles with slopes in the range 1.5-2.0 are put forward for molecular clumps by various studies (see \cite{parmentier:11} and references therein). This means that the matter available for star formation has the same dependency.  In the first case one would end up with the gas mass increasing as $ M_c \propto r^2$ and the second case would result in a dependency  $ M_c \propto r^1$. Assuming a constant star formation efficiency throughout the cluster and using the size-age relation, this corresponds to the cluster mass increasing as $ M_c \propto t^ {2c}$ and $ M_c \propto t^ {c}$, respectively. Generally, one could conclude that for a linear cluster expansion ($c$=1), the gas density distribution in the clump from which these clusters form would have to be of the form  $ M_c \propto r^{1.3}$.

An alternative picture would be that it is not the density distribution that is responsible  for the shape of the observed density-age relation, but the way that star formation proceeds. Here it is assumed that star formation continues in the central area while the star formation front moves outwards in a medium with a $r^{-2}$ gas density dependence. This is a likely scenario, as star formation would probably not immediately stop in the centre but continue until enough massive stars have formed  to blow a cavity into the gas in the central area. For simplicity in Fig.~3, a model  assuming linearly propagating star formation has been calculated (dashed line). In this picture in each shell of thickness $\Delta r$ initially the same amount of star formation $\Delta M$ happens, but the further a shell is from the center the later it starts to produce stars. After its onset the star formation rate slows down by the factor $\alpha_r$ in each of the individual shells as the gas content in these regions decreases.  Assuming that the star formation proceeds over $N$ shells, the total star formation $M_{SF} $ is given by

\be
M_{SF}(N) =  \sum_{i=i}^N \sum_{t=1}^N \alpha_r^{t-i} \Delta M
    =  \sum_{i=i}^N (N-i+1)\alpha_r^{N-i+1} \Delta M.
\ee

 This model fits well with the derived density radius relation shown in Fig.~3 if we choose $\alpha_r(t_{n+1}) = 0.7*\alpha_r(t_n)$.

These simple estimates should be regarded as a first attempt to show how new insights on embedded cluster dynamics might provide clues to the star formation process itself. This should motivate future observations as well as more comprehensive numerical modelling.

\subsection{Gas expulsion}

\begin{figure}
\includegraphics[width=84mm]{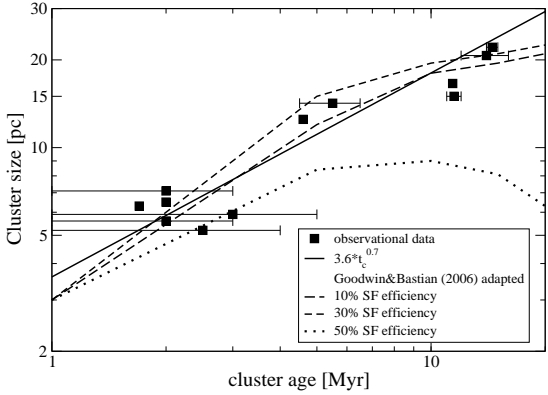}
\caption{Cluster size development as a function of cluster age for exposed clusters. The observational values were taken from Wolff et al. (2007). The simulation data points are from Goodwin \& Bastian (2006). a linear fit in the log-log space is shown as drawn line, scaled versions of the results of Goodwin \& Bastian (2006) for instantaneous gas expulsion with 10\% (short dashed line), 30\%(long dashed line) and 50\%(dotted line) star formation (SF) efficiency
are shown.
\label{fig:mass_radius}}
\end{figure}

As Fig. 2 demonstrates there are two distinct phases in the density development of leaky clusters
- the star formation phase which leads to a density/radius relation as expressed by Eq. 2 and
the cluster expansion phase which is described by Eq. 1. We stress that while the sequence of leaky clusters stems from {\em dynamical evolution } (i.e. post-gas expulsion expansion), the embedded cluster track is the imprint of the  {\em formation process}. The "switch" from one to the other process is the gas expulsion as illustrated by Fig. 3.

Since the cluster expansion and mass loss are the result of the star formation process and the subsequent gas expulsion, it should be possible to draw conclusions about the early history of these clusters from the properties of these two processes. Using results from existing numerical simulations  we can place constraints on how the gas expulsion is likely to proceed in leaky clusters.  

\cite{goodwin:06} happened to simulate a situation similar to that in leaky clusters. Their cluster simulations contained 15 000 stars distributed according to a Plummer modell with a core radius of 1 pc. They investigated the process of instantaneous gas expulsion by assuming virial equilibrium before gas expulsion and varied the star formation (SF) efficiency. Here SF efficiency is defined as $M_ {stars}/(M_ {stars}+M_ {gas})$ with $M_ {stars}$ being the mass of the stars in a cluster and $M_ {gas}$ the gas mass.
Note, that leaky clusters have a somewhat larger number of stars, typically $N \sim$ 20000-40000.  

Fig. 4 shows the radial development of leaky clusters as a function of cluster age for the data used in \cite{pfalzner:09}. Here the drawn line represents Eq. 1 which is basically a linear fit in the log-log space. Note the relatively large errors of 1-2 Myr in cluster age. In addition Fig. 4 shows
the simulation results of \cite{goodwin:06} scaled to the initial cluster parameters relevant here. Such a scaling is possible since the model excludes scattering processes between the cluster stars. It can be seen that a low SF efficiency of 10\% leads to somewhat smaller cluster radii than observed whereas a  high SF efficiency of 50\% would lead only to an increase of cluster radii by a factor of $\sim$ 2.  The best fit is obtained by an instantaneous gas expulsion with a 30\% SF efficiency. In this case the simulated and observed leaky cluster expansion are consistent. Note that the data of the 30\% SF efficiency simulation actually seem a better fit to the observational data than Eq. 1. However, the age determination has such a large error that it is not possible to decide which one is more realistic. Only better observational data and simulations tailored to the actual parameters of leaky clusters will give  an answer to this question. 

It has been suggested that the distinction between starburst and leaky clusters is basically
identical to that of systems remaining bound and becoming unbound after gas expulsion
\citep{portegies:10,gieles:11} and that the leaky clusters are therefore (unbound) associations.
However, taking the above derived SF efficiency of 30\%, the results of \cite{goodwin:06} indicate
that despite the considerable mass loss of these clusters, about 10\%-15\% of the stars can be expected to remain bound and develop into a cluster which lasts $>$ 100Myr. Although much reduced in mass, these bound entities still consist of $\sim$ 1000 stars albeit with a low volume density of the order of 1 \Msun pc$^{-3}$. It is a matter of cluster definition, if such a group of star is still called a cluster.  If one regards boundedness of a sufficient number of stars as prime property of a cluster then it is. By contrast, if one requires a certain volume or surface density  (for a review of the different definitions see \citet{bressert:10}) then it is a cluster at the start of the expansion but can no longer regarded as such at later stages. In either case more detailed investiagtions should be carried out to specifiy the properties of these remaining entities more precisely.

\section{Conclusion}

In this paper we investigated the formation and early development of leaky clusters. This first approach is far from definitive, however, it does provide a consistent picture. We showed that the difference
between leaky and star burst clusters is not mass related but must lie in their early development
(see Fig. 1). Taking our cue from the size/age relation demonstrated for young leaky exposed clusters, we postulated
an analogous relation for the embedded phase.  Based on this concept,

\begin{itemize}
\item 
we examine the star formation history of leaky clusters by determining the cluster radius development as opposed to the traditional examination of the stellar age distribution in single clusters. Here the global properties of a multitude of known embedded clusters is used to make deductions about the star formation history inside individual star clusters.  This is only possible because recent findings strongly suggest that there exists a common developmental history for leaky clusters.
 
\item
the radial cluster development implies that star formation initially proceeds at a low rate and accelerates considerably later on. Spatially this is consistent with star formation in a sphere of increasing volume. 

\item 
the increase of cluster mass with cluster size follows a dependence of the form 
$M_c =a r_{cl}^b $
with $a$=359$\pm$15 and $b$=1.71$\pm$0.07. 
and the cluster density develops as $ \rho \approx 100 r_{cl}^{-1.3} $. 

\item
it has been shown that the observational results of cluster expansion are consistent with a
30\% star formation efficiency followed by an instantaneous gas expulsion.

\end{itemize}

From the existing data, it is not absolutely clear, whether the star formation process is the result of two distinct generations of stars, or a continuous but accelerated star formation. From 
the mass-radius diagram (Fig.~3) the latter seems more likely. 

With these results a picture of the mass and density development of leaky clusters can be constructed over the first 20 Myr of their existance: Leaky clusters start out with several tens of stars being confined within $\sim 0.2-0.3 pc$, grow in mass and cluster membership as $M_c 
\sim 360 r_{cl}^{1.7} $ until they reach their maximum mass of several 10$^4$ \Msun with a size of $\sim$ 3-5 pc. 
Then gas expulsion sets in and the clusters expand rapidly while loosing
most of their mass over the follwing 20 Myr.  
Throughout the cluster formation phase the density (at least in the central region) is so high
that the cluster environment might significantly influence the planet formation process. 

In this paper the focus was on the early evolution of leaky clusters, it would be highly desirable to obtain the equivalent information for starburst clusters. However, at present it is impossible to determine in which way star formation in starburst clusters differs from that in leaky clusters. Fig. 1 implies that the precursors of starburst clusters would form an extension of the starburst cluster sequence towards smaller radii ($<$ 0.1pc) at density of $>$ 10$^{5-7}$ \Msun pc $^{-3}$. Otherwise it would require them to contract and increase in density while simultanuously forming a high number of massive stars. No massive embedded cluster has been definitely classified as a starburst precursor so far possibly with the exception of W43 \citep{motte:03,nguyen:11} This is not surprising as starburst clusters are situated either close to the Galactic centre (8kpc) and/or along spiral arms. At these locations the higher background density makes cluster detection significantly more difficult, so that even most of the exposed massive starburst clusters were only detected recently.



\section{Acknowledgments}

I wish to thank A. Stolte, C. Olczak, G. Parmentier and M. Gieles for very helpful discussions.

\end{document}